# A low phase noise microwave source for high performance CPT Rb atomic clock

Xiaodong Li, Peter Yun✉, Qinglin Li, Bowen Ju, Shaoshao Yu, Qiang Hao, Runchang Du, Feng Xu, Wenbing Li, Yuping Gao and Shougang Zhang

Phase noise of the frequency synthesizer is one of the main limitations to the short-term stability of microwave atomic clocks. In this work, we demonstrated a low-noise, simple-architecture microwave frequency synthesizer for a coherent population trapping (CPT) clock. The synthesizer is mainly composed of a 100 MHz oven controlled crystal oscillator (OCXO), a microwave comb generator and a direct digital synthesizer (DDS). The absolute phase noises of 3.417 GHz signal are measured to be -55 dBc/Hz, -81 dBc/Hz, -111 dBc/Hz and -134 dBc/Hz, respectively, for 1 Hz, 10 Hz, 100 Hz and 1 kHz offset frequencies, which shows only 1 dB deterioration at the second harmonic of the modulation frequency of the atomic clock. The estimated frequency stability of intermodulation effect is 4.7×10$^{-14}$ at 1s averaging time, which is about half order of magnitude lower than that of the state-of-the-art CPT Rb clock. Our work offers an alternative microwave synthesizer for high-performance CPT Rb atomic clock.

*Introduction:* The passive microwave vapour cell atomic clock has been widely used in global navigation satellite systems (GNSS), high-speed communication and other fields due to its excellent performance, simple structure, small size, and low power consumption [1]. Thanks to the advances in semiconductor laser, coherent population trapping (CPT) atomic clocks and laser pumped Rb clocks are widely studied and their shot noise limit has made nearly one order of magnitude improvement. The short-term stability of vapour cell atomic clock less than $5 \times 10^{-13} \tau^{-1/2}$ has been obtained by several laboratories [2-6]. Meanwhile, intensive researches also have been carried out to achieve better long-term stability. To mitigate the temperature-related effects, more compact physics packages and rigid temperature control have been employed in compact vapour cell clocks [7]. To reduce the light shift, the symmetric auto-balanced Ramsey interrogation was applied in a CPT clock, and consequently, a mid-term frequency stability of $2.5 \times 10^{-15}$ at $10^4$ s was demonstrated [8].

As is well known, phase noise of the microwave interrogation signal is one of the main contributions to the short-term stability of microwave clocks via Dick effect for a pulsed clock or intermodulation effect for a continuously operated clock [9-12]. Taking the latter clock as an example, the technique of microwave frequency modulation and synchronous demodulation is utilized to tightly lock the local resonator to the atomic resonant signal. When the microwave signal is modulated with a low-frequency $f_m$, the phase noise at offset frequencies which equal the even multiples of $f_m$ translate frequency fluctuations into the probe signals of the servo loop. The numerical evaluation to the frequency stability $\sigma_y(\tau)$ due to the intermodulation effect with sine wave modulation is given by [13]

$$\sigma_y^2(\tau) = \sum_{n=1}^{\infty} C_{2n}^2 S_\varphi(2nf_m) \tau^{-1} \quad (1)$$

where $S_\varphi(2nf_m)$ are the power spectral densities (PSD) of phase fluctuations, $C_{2n}$ are the coefficients for different harmonics. For vapor cell atomic clocks, the modulation frequency is generally at hundreds Hz level and the contributions of intermodulation effect are dominated by the low-order harmonics. Therefore, the phase noises at 100 Hz - 100 kHz are usually taken into account in vapour cell clocks.

A low-noise microwave synthesizer is indispensable for a high-performance atomic clock. The portable microwave synthesizers based on nonlinear transmission line (NLTL) and sub-sampling phase lock loop have been successfully applied in compact clocks [14, 15]. However, in these studies, the dielectric resonant oscillator (DRO) has to be phase-locked to oven controlled crystal oscillator (OCXO) because the phase noise performance of free-running DRO is quite poor. Francois *et al.* demonstrated a quasi-ideal microwave frequency synthesizer, without degradation of the OCXO initial phase noise. However, this synthesizer used a non-negligible number of components, with a frequency key-step at 1.6 GHz and additional converters, filters and amplifiers [16]. To further simplify the synthesizer and close to the performance of ideal multiplier, the scheme with a frequency multiplier based on a step recovery diode (SRD) and a DDS is presented here.

*Design scheme:* The frequency synthesizer refers to a 100 MHz OCXO and outputs a 3.417 GHz microwave signal by mixing the 3.5 GHz signal from a comb generator and the 83 MHz signal from a DDS. We call it as "3.5 GHz-83 MHz" solution. Similarly, there are "3.4 GHz+17 MHz", "3.6 GHz-183 MHz" and other solutions, whereas the "3.5 GHz-83 MHz" is chosen due to the following considerations. Firstly, the narrow bandwidth band-pass filter (BPF), which is needed to filter out the 3.4 GHz signal from the 3.417 GHz signal after the mixing, is costly and bulky. Secondly, with the same reference frequency, the phase noise of DDS output signal becomes worse as the output frequency increased [17].

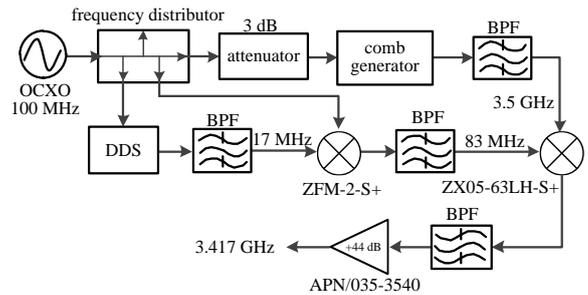

**Fig.1** *The architecture of 3.417 GHz microwave synthesizer.*

As shown in Fig.1, the OCXO named AXIOM5050ULN with ultra-low phase noise is chosen as the local oscillator for the microwave frequency synthesizer. The output signal of the local oscillator is divided into four arms via a frequency distributor (Sync-tech, STMFP4-1-120), and the output power of each arm is about +13 dBm. The first arm is connected to the comb generator (CETC, HGC101-9) and then generates the high-order harmonics of the 100 MHz sinusoidal signal (see Fig. 2(a)), with the help of the BPF (Yun Micro Electronics, RB3500-40-8R5CCS), the harmonics of 100 MHz signal except 3.5 GHz are fully suppressed and the output power is about -27dBm (see Fig. 2(b)).The second arm is used as the low phase noise reference for the DDS (ADI,AD9912), which outputs a 17 MHz low-frequency signal, and harmonics of DDS output signal can be well suppressed with a BPF (Yun Micro Electronics, YMB17-2-8R4CCS). Then the 17 MHz signal and the 100 MHz signal from the third arm go into a mixer (Mini-Circuits, ZFM-2-S+) and a BPF (Yun Micro Electronics, YMB83-8-5C1AS) to produce a pure 83 MHz signal. Finally, the 3.5 GHz and 83 MHz signals are mixed (Mini-Circuits, ZX05-63LH-S+) and band-pass filtered (Yun Micro Electronics, RB3417-40-8R5CCS) to generate a 3.417 GHz microwave signal. This signal is amplified to about +16 dBm by a power amplifier (CTT, APN/035-3540) enabling the subsequent measurement and atomic clock application. In addition, the components in the chain are all commercially available.

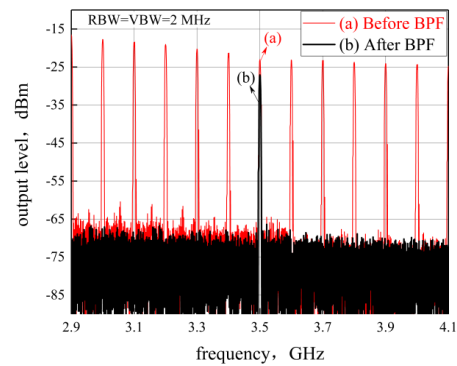

**Fig.2** *The output signal spectrum of the comb generator in the 3 GHz-4 GHz range before BPF and after BPF.*

From the above description, we can know that the comb generator is the core component of the entire frequency synthesis chain due to its phase noise performance dominating that of the output microwave signal. The comb generator works as follows: after amplified by the preamplifier inside it, the input signal is power-sufficient to drive the SRD to generate the high-order harmonics. Precisely because the preamplifier is power-dependant, the phase noise of 3.5 GHz signal depended on the input power of the comb generator.

The relationship between the phase noise performance of the comb generator output signal and the 100 MHz RF input power is shown in Fig.3. When the input power is about +10dBm, the phase noise (with offset frequency $\geq$ 500 Hz) of the 3.5 GHz signal is the best. The measurement of the phase noise is implemented by a high-performance cross-correlation signal source analyzer (APPH20G), which can meet the requirements in this work.

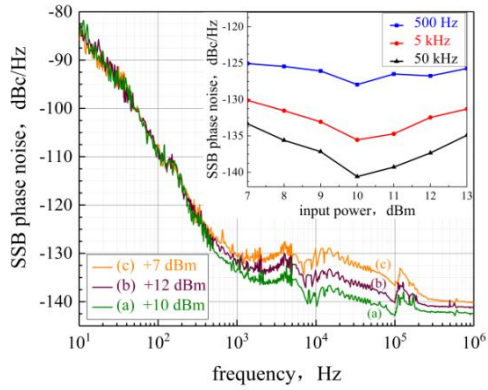

**Fig.3** *Absolute phase noise of 3.5 GHz signal under different input powers, e.g. 7 dBm, 10 dBm and 12 dBm. The inset shows the phase noise performances of 3.5 GHz signal versus the input powers of 100 MHz RF at the offset frequencies of 500 Hz, 5 kHz and 50 kHz respectively.*

It can be seen from Fig.1 that the 100 MHz output signal of the OCXO splits into several arms before used, thus a frequency distributor is required and its phase noise performance is also very crucial. We have measured several distributors from different suppliers and compared them in Fig.4, the customized one (STMFP4-1-120) has slightly better (about 2 dB) phase noise performance for $f < 500$ Hz than the others, and much better (about 9 dB) for $f > 500$ Hz. Thus STMFP4-1-120 is used in the frequency synthesis chain.

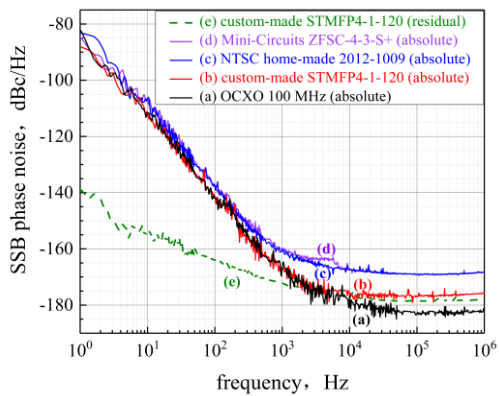

**Fig.4** *Absolute phase noise of (a) the 100 MHz signal of the OCXO and the 100 MHz output signals of the three frequency distributors, (b) the customized distributor STMFP4-1-120, (c) the home-made one, (d) the commercial Component Mini-Circuits ZFSC-4-3-S+. And (e) residual phase noise of the customized distributor STMFP4-1-120.*

We also report the residual phase noise of the frequency distributor STMFP4-1-120 in Fig.4, and found the absolute phase noise (with offset frequency >10 kHz) of the OCXO is deteriorated by the frequency distributor about 5 dB. In order to keep the later analysis intuitive, we calculated the theoretical phase noise of the 3.417 GHz signal from the frequency distributor output according to the deterioration rule of frequency multiplication *i.e.* 20lg*N* (dB), where *N* is the frequency multiplication factor.

*Experimental results:* Based on the work above, we measured the absolute and residual phase noise performances of the key signals and components of the frequency synthesis chain as shown in Fig.5, and the theoretical phase noise of the 3.417 GHz signal is added for reference. From curve (e) and (f), we find that the phase noise of the 3.417 GHz signal is basically identical with performance of the 3.5 GHz signal. In the offset range of 1 Hz-1 kHz, the phase noises of the microwave signal, only 1-2 dB higher than the theoretical values, are mainly dominated by that of the OCXO. It is worth noticing that the phase noise is -117dBc/Hz at 200 Hz, the second harmonic of the microwave modulation frequency, which is less than 1 dB of degradation compared with the theoretical value. For 1 kHz $< f <$ 9 kHz, the phase noises of the 3.417 GHz signal are deteriorated by about 4-5 dB compared with the theoretical prediction, which is caused by the phase noises of both the OCXO and the comb generator. Especially at the offset frequency of about 10 kHz, the severest deterioration of 7 dB occurred. The phase noise in the 9 kHz-90 kHz is limited by the residual phase noise of the comb generator and the deterioration essentially originates from the poor phase noise floor of the SRD. For offset frequencies above 90 kHz, besides the residual phase noise of the comb generator, that of the frequency distributor is not negligible for the deterioration of about 5 dB. Moreover, the phase noise of the 83 MHz signal from the DDS does not degrade the performance of the 3.417 GHz signal, which benefits from the ultra-low phase noise 100 MHz clock source and the low residual phase noise of the DDS (see curve (a) in Fig. 5).

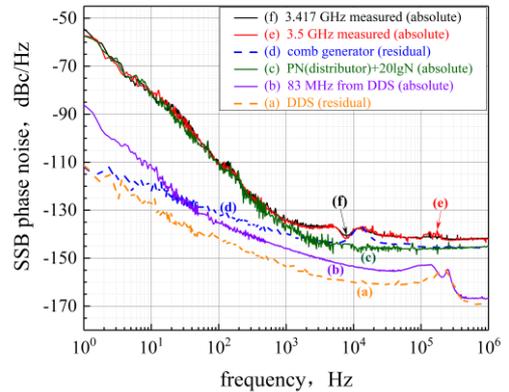

**Fig.5** *Absolute phase noise performances of the key signals of the synthesis chain, (b) the 83 MHz output signal from the DDS, (c) the theoretical phase noise of 3.417 GHz signal from the 100 MHz frequency distributor according to the deterioration rule, (e) the 3.5 GHz output signal of the comb generator, (f) the finally measured 3.417 GHz signal. And residual phase noise of the key components, (a) the DDS of 83 MHz output signal, (d) the 3.5 GHz comb generator.*

More specifically, the absolute phase noise performances of this work are quantitatively compared with that at 3.417 GHz in several previous works as listed in Table 1. The performances in this work are a little better than those of the Ref. [14] and [15]. The frequency synthesizer in Ref. [16] is the best in terms of phase noise in these related works at the cost of more multipliers, dividers and filters than our scheme. Furthermore, the synthesizer in Ref. [18] uses a similar approach utilizing a SRD, however, the residual phase noise of the synthesis chain in our scheme is better than the one in Ref. [18], which is mainly caused by the different distributors and the SRDs used, and the noise floor of analyzer in Ref. [18] may also set a limitation for the measurement of the residual phase noise of the synthesis chain.

**Table 1:** The phase noises at 3.417 GHz of several related works.

| Offset frequency (Hz) | Ref. [14] (dBc/Hz) | Ref. [15] (dBc/Hz) | Ref. [16] (dBc/Hz) | This work (dBc/Hz) |
|---|---|---|---|---|
| 100 | -108.5 | -109.7 | -111 | -110.7 |
| 1k | -125.5 | -127.2 | -136 | -134.3 |
| 10k | -137.7 | -132.7 | -144 | -138.3 |
| 100k | -138.3 | -130.3 | -146 | -141.7 |

With the absolute phase noises of the 3.417 GHz microwave obtained with our method and estimated from the formula (1), the microwave's contribution to the atomic clock frequency stability through intermodulation effect is $4.7 \times 10^{-14}$ at 1 s averaging time, which is about half order of magnitude smaller than the state-of-the-art frequency stability of the CPT clock. We have also measured the residual frequency stability of the 3.417 GHz microwave chain with the similar method in Ref. [19]. The 100 MHz OCXO splits into three arms, the first arm drives the 3.417 GHz signal chains as shown in Fig. 1, the second arm drives the 3.427 GHz signal with the same architecture. Then the two microwave signals are mixed to generate a beat note at 10 MHz, the beat-note signal is then compared with the third arm of the 100 MHz signal using the frequency counter (VREMYA-CH, VCH-323). The residual frequency stability of the 3.417 GHz microwave chain is measured to be $7.1 \times 10^{-15}$ and $1.1 \times 10^{-16}$ for averaging times of 1 s and 1000 s, which is about 20 times lower than that of the state-of-the-art performance of the compact microwave clock.

*Conclusion*: We have demonstrated a low phase-noise 3.417 GHz microwave frequency synthesizer with the simple architecture and the high spectral purity. The absolute phase noises are -111 dBc/Hz, -134 dBc/Hz, -138 dBc/Hz and -142 dBc/Hz at 100 Hz, 1 kHz, 10 kHz and 100 kHz offset frequencies respectively. For the CPT atomic clock, the short-term frequency stability limited by the absolute phase noise via the intermodulation effect is theoretically estimated to be $4.7 \times 10^{-14}$ at 1 s averaging time. So we come to the conclusion that the microwave frequency synthesizer in this work can meet the requirements of the high-performance CPT atomic clocks. Due to the limitations of the residual phase noise of the frequency distributor and the comb generator, the phase noise performance of the microwave frequency synthesizer could be improved further at the offset frequencies larger than 2 kHz, which will be investigated in the future studies.

*Acknowledgments:* This work was funded by the National Natural Science Foundation of China (grant no. U1731132).

One or more of the Figures in this Letter are available in colour online.

Xiaodong Li, Peter Yun, Qinglin Li, Bowen Ju, Shaoshao Yu, Qiang Hao, Yuping Gao and Shougang Zhang (*National Time Service Center, Chinese Academy of Sciences, Xi`an 710600, Shaanxi, People's Republic of China*)

✉ E-mail: yunenxue@ntsc.ac.cn

Runchang Du (*Chengdu Spaceon Technology Co., Ltd., Chengdu 611731, Sichuan, People's Republic of China*)

Feng Xu (*China Academy of Space Technology (Xi'an), Xi'an 710600, Shaanxi, People's Republic of China*)

Wenbing Li (*Hubei Key Laboratory of Gravitation and Quantum Physics, PGMF and School of Physics, Huazhong University of Science and Technology, Wuhan 430074,Hubei, People's Republic of China*)

Qinglin Li, Bowen Ju and Shaoshao Yu: also with University of Chinese Academy of Sciences, Beijing 100049, People's Republic of China

Wenbing Li: also with PRISMA Cluster of Excellence, Johannes Gutenberg Universität , Mainz 55099, Federal Republic of Germany